\documentstyle[12pt,aaspp4]{article}
\tightenlines

\begin{document}

\begin{center}
Accepted for publication in {\it The Astronomical Journal}
\end{center}

\vskip 24pt

\title{A Deep Multicolor Survey V: \\ 
The M Dwarf Luminosity Function}

\vskip 24pt

\author{Paul Martini and Patrick S. Osmer}

\vskip 12pt

\affil{Department of Astronomy, 174 W. 18th Ave., Ohio State University, \\
Columbus, OH 43210 \\
martini,posmer@astronomy.ohio-state.edu}

\vskip 48pt

\centerline
{\bf Abstract}

We present a study of M dwarfs discovered in a large area, multicolor 
survey. We employ a combination of morphological and color criteria to 
select M dwarfs to a limiting magnitude in $V$ of 22, the deepest such
ground-based survey for M dwarfs to date. We solve for the vertical disk 
stellar density law and use the resulting parameters to derive the M dwarf 
luminosity and mass functions from this sample. We find the stellar luminosity 
function peaks at $M_V \sim 12$ and declines thereafter. Our derived mass 
function for stars with $M < 0.6 M_{\odot}$ is inconsistent with a 
Salpeter function at the $3\sigma$ level; instead, we find the mass function
is relatively flat for $0.6 M_{\odot} > M > 0.1 M_{\odot}$. 


\clearpage

\section{Introduction}

The study of M dwarfs in the disk of our galaxy has applications to many
areas of current research, including the determination of the mass of the
disk, its star formation history, and the nature of the sources of 
observed microlensing events (e.g. Bahcall et al.\ 1994). The shape of the 
stellar luminosity function has been measured using a variety of techniques, 
including large area, bright, ground-based surveys (e.g. Stobie, Ishida, 
\& Peacock 1989; Reid, Hawley, \& Gizis 1995; Reid \& Gizis 1997), 
deeper ground-based
surveys with complementary spectroscopy (Reid et al.\ 1996), and 
space-based surveys performed with the Hubble Space Telescope (Gould, Bahcall, 
\& Flynn 1996; 1997). These have all shown that the shape of the stellar 
luminosity function peaks at $M_V \sim 12$ and declines thereafter to at 
least $M_V \sim 15$. 

Each of these techniques has particular strengths. Large-area, ground-based
surveys to relatively bright magnitude limits 
are particularly suited to determining the bright end of the M dwarf luminosity
function in the solar neighborhood, where the effects of the space-density 
gradient in the M dwarf population can be neglected. Deeper, ground-based
surveys may be readily used to obtain a large sample of faint M dwarfs. These 
deeper surveys are generally limited to much smaller areas and can suffer from 
contamination from faint, red galaxies. Deep, ground-based M dwarf counts 
therefore require spectroscopy to eliminate this contamination and to 
accurately measure the stellar luminosity function. Finally, star counts
with HST can reach to the faintest limiting magnitudes by distinguishing 
stars from galaxies with the excellent angular resolution of HST. 
Deep, space-based surveys preferentially sample M dwarfs higher above 
the plane than ground-based, larger-area surveys to brighter limiting 
magnitudes. They therefore probe both the shape of the stellar luminosity 
function and the space density of the M dwarf population, though they are 
not as sensitive to the brighter M dwarfs as they probe smaller areas. 
These approaches therefore preferentially sample different regions of the 
phase space needed to determine both the M dwarf luminosity function 
and the space density of this population as a function of height above the 
plane and distance from the galactic center. 

In the present work we derive the shape of the vertical stellar density 
distribution of M dwarfs and use this to compute their luminosity and mass 
functions using the deep multicolor survey (DMS) of Hall et al.\ (1996a). 
This survey has already been successfully used to study quasars 
(Hall et al.\ 1996b; Kennefick et al.\ 1997) and the luminosity function of 
galaxies (Liu et al.\ 1996). In this contribution we search the DMS stellar 
catalog (Osmer et al.\ 1998) for M dwarfs down to faint limiting magnitudes. 
However, in contrast to previous surveys, we primarily employ extensive 
multicolor information, rather than spectroscopy, to separate M dwarfs from 
faint, red galaxies. We do, however, make use of the spectroscopic 
identifications obtained of selected M dwarfs as part of the search for 
quasars as a check of the efficiency of our multicolor selection criteria. Our 
final catalog of M dwarfs extends to $V = 22$ and contains a total of 1575 M 
dwarfs, including 499 in bins over the range $12 \leq M_V \leq 15$.

\section{Observations and Sample Selection}

The multicolor survey fields obtained by Hall et al.\ (1996a) image 
six different high galactic latitude $(|b_{II}| > 35^{\circ})$ fields in six 
filters and subtends a total of $0.83\;\Box^{\circ}$. These six filters are 
Johnson $UBV$ and three special red filters, $R'$, $I_{75}$, and $I_{86}$. 
The $R'$ filter is somewhat narrower than a Harris $R$ or Cousins $R$ and
has less of a red tail. The $I_{75}$ and $I_{86}$ filters are narrow-band
$I$ filters centered at 7500 \AA\ and 8600 \AA, respectively. Hall et al.\ 
(1996a) used FOCAS to catalog all of the stellar objects in these 
fields in order to 
search for quasars (Hall et al.\ 1996b; Kennefick et al.\ 1997). An object was 
declared stellar provided it was classified as a star or fuzzy star in more
than one half of the filters in which it was detected, and each object was 
required to be detected in at least three filters for inclusion in the 
final stellar catalog. We note that this morphological criterion is extremely 
lenient because the goal was not to exclude AGNs, which may be slightly 
resolved; we do not expect it to 
have missed any stellar objects. The resulting catalog has $5\sigma$ limiting
magnitudes between 22.1 and 23.8 for these six filters. 
We refer the reader to Hall et al.\ (1996a)
for a more detailed explanation of the filters, catalog construction, and 
limiting magnitudes summarized here. We summarize the field size and 
galactic coordinates for each field in table 1. 

Reid et al.\ (1996), using LRIS on the Keck telescope, find $R = 22.5$ 
to be the limit for star-galaxy separation in their good-seeing ($0.5''$),  
ground-based data with a 10\% contamination due to
faint galaxies. They find the galaxy contamination to be 30\% by $R = 24.5$. 
The data discussed here had typical seeing
between $1 - 2''$ FWHM and we therefore use a combination of morphological 
and color criteria in order to extend the limiting magnitude of accurate 
star/galaxy separation as faint as possible. However, the three reddest 
filters used in this 
investigation are nonstandard and therefore measurements of M dwarf colors with 
these filters do not exist in the literature. Fortunately, the $R'$ filter 
has nearly the same effective wavelength as 
the KPNO Harris $R$ (which has the same transmission profile as the 
Cousins $R$; hereafter $R_C$), though slightly narrower and with less of a red 
tail. The $R'$ magnitudes quoted by Hall et al.\ (1996a) are therefore 
equivalent to $R_C$ except perhaps for the reddest objects, where the $V-R'$ 
color may be slightly bluer than $V-R_C$ due to the weaker
red tail of the $R'$ filter. 

To measure the magnitude of this potential 
color term, we used the STSDAS SYNPHOT package to convolve a sample of 
late-type stars from the Bruzual-Persson-Gunn-Stryker Spectrophotometry Atlas 
with the $V, R_C,$ and $R'$ filters. We found the color term would be no 
greater than 0.02 in $V-R_C$ for the reddest M dwarfs. As the standard stars 
used in the photometric calibration included stars as red as $V-R_C \sim 1.3$ 
(Hall 1998 {\em private communication}), while the M dwarfs studied here have 
$0.9 < V-R < 1.6$, this is not surprising. It is also interesting to note the 
direction of this effect. The transformation from $R'$ to $R_C$ makes red 
objects brighter in the $R_C$ filter. This corresponds to redder $V-R_C$ and 
hence later spectral type and larger $M_V$. Thus any color term for the 
reddest objects would make them intrinsically fainter and therefore closer. 
Our assumption that $R'$ is equivalent to $R_C$ may therefore slightly 
increase the space density of the faintest M dwarfs. 

The transmission profiles and narrow widths of the 
$I_{75}$ and $I_{86}$ filters are significantly different from Cousins
$I$. This region of the spectral energy distribution of M stars
is also quite strongly affected by the presence of molecular absorption. 
The combination of these two factors introduces significant uncertainties
into a transformation of $I_{75}$ and $I_{86}$ to Cousins $I$ and we 
therefore did not perform this transformation; rather, this drove us to 
use the $V-R_C$ color to estimate absolute magnitudes. 
 
Ground-based surveys have been able to separate stars from galaxies to 
$R \sim 19$ without spectroscopic follow-up (Kirkpatrick et al.\ 1994). 
To increase the limiting magnitude to which we could successfully discriminate 
M dwarfs from small, red galaxies, we employed the 5 colors of the Hall
et al.\ (1996a) dataset to remove nonstellar objects. These filters 
effectively provide complete wavelength coverage from the atmospheric cutoff 
to 9000 \AA. We expect stars and galaxies to separate in this 
5-dimensional color space. As we are searching for very red objects, however, 
requiring a detection in $U$ or even $B$ will significantly decrease the total 
number of objects in the sample extracted from the catalog. We therefore 
developed the following set of color and detection criteria for extracting a 
catalog of potential M dwarfs. 

Our first cut was to require detection in 
$V$, $R_C$, the two $I$ filters, and a $V$ magnitude less than 23.5, the 
$5\sigma$ limiting magnitude in $V$ (though we refine this limit below). 
The purpose of this cut was to trim the size of the stellar catalog to 
objects with good signal-to-noise ratio photometry in the red filters. 
We then employed a second, much more discriminating cut to take advantage of 
our five color information. We used the M dwarf color information tabulated by 
Leggett (1992) to set the following blue limits for a given star to be included 
in the M dwarf catalog, requiring that the star be within 3$\sigma$ of an 
M0 dwarf (or redder) in each color.

\begin{equation}
\begin{array}{c}
U - B + 3 \sigma_{U-B} > 1.15 \\
B - V + 3 \sigma_{B-V} > 1.35 \\
V - R_C + 3 \sigma_{V-R_C} > 0.90 \\
\end{array}
\end{equation}

As stated above, we did not require that a given star be detected in
the $U$ and $B$ filters. If the 1$\sigma$ photometric uncertainty for $U-B$ 
and/or $B-V$ color was greater than 0.333, corresponding to a nondetection 
at the 3$\sigma$ level, we did not include this color criterion. Though $U$
and $B$ data was not available for every star in the sample, these filters
were nevertheless useful for removing anomalously blue objects from our 
M dwarf sample. A blue $U-B$ color, for example, caused us to reject 290 
stellar objects that we would have otherwise included in this sample and a
combination of blue $U-B$ and $B-V$ or $B-V$ alone caused us to reject an 
additional 120 stellar objects. 
We did not employ color selection criteria based upon the two $I$ filters. 
A quick visual inspection of color-color diagrams
including these two bands (see figure 1), however, does not reveal 
a significant population of blue ($R_C - I < 0$) outliers. 

As an independent check of the efficiency of our color selection technique, 
we searched our final catalog for the M dwarfs `serendipitously' discovered in 
the course of the spectroscopic follow-up to search for faint quasars. These 
spectra, and the entire stellar catalog, are described by Osmer et al.\ (1998) 
and include 45 definite M dwarfs with $V < 22$. Our final M dwarf catalog
includes 37 of these objects, or 82 \%. The 8 stars excluded by our 
search criteria were all eliminated due to their anomalous $U$ and $B$ 
brightness, suggesting they may have faint, blue companions. As the presence
of an ultraviolet excess is one of the selection criteria for quasars, it
is not surprising that some of the M dwarfs observed spectroscopically are 
associated with 
excess UV emission. In fact, the sample of M dwarfs with spectroscopic
confirmation is biased towards stars with unusual $UV$ colors and this sample
cannot be used to characterize the true efficiency of the color selection 
criteria for M dwarfs. Rather, the 82 \% completeness quoted above can 
only be considered a strong lower limit to the true completeness of this 
color-selected M dwarf sample.

\section{Color-Magnitude Relation}

We have employed the $V-R_C$ color to estimate $M_V$ for the M dwarfs in our 
sample. We derived the $M_V,V-R_C$ relation as follows: we first collected all 
of the stars with parallax data and $0.9 < V-R_C < 1.65$ (no redder stars are 
contained in our sample) from the work of Leggett (1992). We then 
cross-referenced these stars with the Hipparcos catalog (ESA 1997) and 
substituted the Hipparcos parallax measurements when they were more precise. 
As a final cut, we removed unresolved, known double systems and stars with 
uncertainties in $M_V$ greater than 0.2 mag. Our resulting sample of 177 
stars, along with our fit, is shown in figure 2. The best-fit color-magnitude
relation is: 

\begin{equation}
M_V = - 5.50 + 20.20 \; (V - R_C) - 4.48 \; (V - R_C)^2. 
\end{equation}

\noindent
The unbiased mean deviation for the fit is $\sigma_{M_V} \sim 0.52$. 
We differentiate between members of the young ({\it crosses}), intermediate 
({\it triangles}), and halo ({\it circles}) populations based on the 
kinematic 
data provided by Leggett (1992). We plot only error bars if the kinematic 
population membership was not available. We represent different kinematic 
populations with different symbols as kinematic population type is known to 
correlate with metallicity (e.g. Carney, Latham, \& Laird 1990; Leggett 1992), 
which in turn may affect the color-magnitude relation because of the impact 
of the TiO bands on the $V$ and $R_C$ bands. In our analysis, we only find 
evidence for a different color-magnitude relation in the halo population, which 
comprises approximately 12\% of the calibration sample. We find the halo stars
are approximately 0.5 mag fainter for a given $V-R_C$ color. As metal-poor 
stars are expected to be not only fainter for a given $V-R_C$, but may also
contribute a relatively larger fraction of the stars observed further from the
galactic plane, they may introduce a systematic effect. To see if our data 
shows evidence for a metallicity 
gradient in the M dwarf population, we examined the distribution of $B-V$ 
color at a given $V-R_C$ color as a function of height above the plane. This 
analysis did not reveal a significant color change with height above the plane 
and we therefore conclude from this result, and the relative scarcity of
low-metallicity stars, that our results will not be strongly affected by a 
vertical metallicity gradient in the galactic disk or a low-metallicity 
component. 

We note that most investigators (e.g. Stobie et al.\ 1989; Reid 1991; Gould et 
al.\ 1996) employ a linear relation between $V-I$ and $M_{V}$ to estimate 
stellar luminosities. Reid \& Gizis (1997) have shown, however, that 
there is a change in the slope of this relation at $V-I \sim 2.9$. They 
attribute much of the current disagreement on the shape of the stellar 
mass function to the use of an inexact $M_V$, $V-I$ relation.

\section{Stellar Density Distribution}

The next step in our analysis was to use the calculated $M_V$ and the 
method of photometric parallax to estimate the perpendicular distance of 
our sample objects from the galactic plane and their distance from 
the galactic center. Stobie et al.\ (1989), in a study 
of the stellar luminosity function within 125 pc and in the direction 
of the north galactic pole, employed a single 
exponential distribution with a scale height of 325 pc (Yoshii et al.\ 1987) to 
estimate the total effective volume (which they refer to as the generalized 
volume) sampled by their survey for the computation of the stellar luminosity 
function. Gould et al.\ (1996; 1997) have used the high angular resolution of 
the Hubble Space Telescope to discriminate M dwarfs from galaxies many scale 
heights above the plane. Gould et al.\ (1996; 1997) used the method of 
maximum likelihood to model the vertical and radial distribution of the 
M dwarfs in their sample and found the vertical stellar density to be 
best fit by a two-component model with a sech$^2$ term and an exponential term, 
along with a parameter $\alpha$ to describe the relative contribution of 
each of these terms. Gould et al.\ also fit a double exponential model to their
data, but found this did not provide as good agreement with their data as the 
sech$^2 +$ exponential model. 

Our sample includes a large number of stars within a kiloparsec of the 
galactic plane. As we have a significantly larger number of stars in this range 
than the HST sample of Gould et al.\ (1996; 1997), we have reexamined the 
functional form of the vertical 
stellar density distribution. Our modeling procedure was as follows: First, we
selected all of the stars in our four brightest luminosity bins, $M_V = 9, 10, 
11,$ and $12$. We then separated out all stars within a kiloparsec of the 
galactic plane and split these stars into 100 parsec bins. This set of 
4 luminosity bins and 10 spatial bins was compared to the number of stars 
predicted by different models with a $\chi^2$ goodness-of-fit estimator. 
We used the downhill simplex method of Nelder \& Mead (1965) as described
by Press et al.\ (1992) to test three models for the vertical stellar density 
distribution: a single exponential, a double exponential, and a sech$^2 +$ 
exponential model. We described the radial component in these models
with the parameterization of Gould et al.\ (1997). Our search of the resulting 
5 or 7 dimensional parameter space found that the two-component sech$^2 +$ 
exponential model provided the best fit to our data. We list the best-fit 
parameters of this model in table 2. Including stars further above the plane 
or adding more luminosity bins to the model did not significantly change the 
best-fit parameters. Our resulting expression for the vertical stellar density 
distribution is: 

\begin{equation}
\nu[z] = 0.79 \; {\rm sech}^2 \left( \frac{z}{340} \right) + 0.21 \; {\rm exp} \left( \frac{-|z|}{550} \right).
\end{equation}

\noindent
We note that the parameters we derived for this model agree with the parameters
derived by Gould et al.\ (1997) to within our uncertainties. After obtaining 
the above solution for the vertical distribution of stars, we attempted to 
solve for the radial distribution with a single exponential fit of variable 
scale length. Our data were not, however, able to provide reasonable 
constraints on the disk scale length. We have therefore used the 
parameterization of the radial stellar density distribution derived by 
Gould et al. (1997): 

\begin{equation}
f[r] = {\rm exp} \left( \frac{8000 - r}{2920} \right).
\end{equation}

\section{Analysis}

To calculate the stellar luminosity function, we used the concept of an 
effective volume to compute the total volume within which a star of 
a given $M_V$ could have been detected. The effective volume is defined as:

\begin{equation}
v_{eff}(M_{V}) = \Omega \int_{d_{min}}^{d} dl \, l^2 \nu[z] f[r]
\end{equation}

\noindent
where $\Omega$ is the angular size of the field in steradians, $\nu[z]$ and 
$f[r]$ are as defined above, and $l$ is the line element over which the star 
could have been detected from $d_{min}$, defined by the saturation limit 
in $V$, to $d$, derived from the distance modulus. 

In order to refine the limiting magnitude of the survey, we used 
the fact that the stellar distribution should be uniform in 
effective volume space. We therefore computed the effective volume for 
each object from the saturation limit to its distance and divided 
this quantity by the effective volume within which the object could 
have been 
detected for a given limiting magnitude in $V$. If the survey is uncontaminated
by small, red galaxies 
down to a given limiting magnitude, the cumulative distribution of 
$v_{eff}/{\rm max}\{v_{eff}\}$ should be linear. If the survey suffers from 
an excess of objects near the limiting magnitude, then the cumulative
distribution will appear to have positive curvature. Similarly, 
if the distribution is incomplete, then the cumulative distribution will
appear to have negative curvature. Finally, a cumulative 
distribution that appears best fit by a high order polynomial would
suggest that the model for the stellar distribution is incorrect. We 
computed the cumulative distribution for magnitude limits from $V = 21$ 
to $V = 24$. From this analysis, we conclude that there is 
insignificant contamination for $V \leq 22$, while for $V > 22$ 
the contamination of the stellar catalog quickly becomes significant. 

A second test to determine the limiting $V$ magnitude at which the 
contamination of the M dwarf catalog begins to become severe is to use the 
$V/V_{max}$ test of Schmidt (1968) as modified by Avni \& Bahcall (1980). 
The $V/V_{max}$ test was applied by Reid et al.\ (1995) on the true space
volume to estimate the completeness limit of a relatively
nearby sample of M dwarfs from a preliminary version of the Gliese \& Jahreiss
(1991) {\it Third Catalog of Nearby Stars} as a function of $M_V$.
Due to the density gradients in the stellar distribution, we calculated 
$V/V_{max}$ with the effective volume, rather than the true space volume. 
Computing $V/V_{max}$ for
different limiting magnitudes is an equivalent to testing for 
contamination or incompleteness at the faint end. If $V/V_{max}$ is 
greater than 0.5, this suggests the survey suffers from contamination 
of nonstellar objects at the faint end. Similarly, values less than 0.5
suggest incompleteness at the faint end. This method, however, is not as 
sensitive to errors in the assumed stellar distribution. Table 3 shows the 
results of the $V/V_{max}$ calculations, where $V/V_{max}$ is consistent with 
0.5 within the $3 \sigma$ statistical fluctuations to approximately $V=22$.

\section{Results}

The luminosity function, $\phi(M_V)$ for a given absolute magnitude bin is: 

\begin{equation}
\phi(M_V) = \sum_{i=1}^{N} \frac{1}{{\rm max}\{v_{eff,i}(M_V)\}} 
\end{equation}

\noindent
where $N$ is the number of stars in bin $M_V$. The resulting luminosity 
function is shown in figure 3 for unit magnitude bins centered at 
integer magnitudes ({\it open circles}). We list the numerical values of
both the luminosity and mass function in table 4. 

The bright end ($M_V < 12$) of the luminosity function is in excellent 
agreement with other recent measurements of the M dwarf luminosity 
function (Stobie et al.\ 1989; Reid \& Gizis 1997; Gould et al.\  1997). 
The faint end of the luminosity function, however, has been significantly 
more controversial due to the selection effects and small 
number statistics which complicate an accurate census of the faintest M dwarfs. 
Stobie et al.\  (1989) measured the luminosity function down to $M_V$ of 16 in 
a large area ($21.46\;\Box^{\circ}$) survey of the north galactic pole and 
found the luminosity function peaks at $M_V = 12$ and decreases over 
the range $12 < M_V < 15$, even when the effects of binary/multiple systems
are taken into account. Reid \& Gizis (1997) found a similar turn down in the 
their rederivation (using their new ($M_V$, $V-I$) relation) of the luminosity 
function from photometric parallax for $M_V = 12$ to 15, though with an 
increase again for the very faintest dwarfs. We note, however, that the 
increase in the photometric luminosity function for $M_V > 15$ has a very low
statistical weight. Gould et al.\ (1997) also find the luminosity function 
peaks at $M_V = 12$, though they also see some evidence for an increase in the 
stellar number density for the faintest ($M_V > 16$) M dwarfs. In agreement 
with these investigators, we find the M dwarf luminosity function turns over at 
$M_V \sim 12$ and declines to at least $M_V = 15$, the limit of our sample. 
The number density we measure at the faint end, shown in figure 3, is 
greater than that of Gould et al.\ (1997; crosses with error bars) and Stobie 
et al.\ (1989; filled triangles), and slightly less than that of Reid \& Gizis 
(1997; filled circles). 

Including M dwarfs in binary systems introduces two competing effects. 
Unresolved, red binaries can artificially increase the luminosity function
by scattering stars into the sample (Reid et al.\  1995), leading to an 
underestimated distance and correspondingly overestimated space density. 
In contrast, unobserved M dwarf secondaries are missed, leading
to an underestimation of their true space density. Reid \& Gizis (1997) 
used their measurement of the binary fraction within 8 pc to model the
effect of unresolved binaries on the luminosity function.
They find that the luminosity function derived via the technique of 
photometric parallax is in general agreement with the true luminosity 
function used as input to their model. 

To convert our luminosity function into a mass function, we adopted the 
empirical mass$-M_V$ relation of Henry \& McCarthy (1993). The resulting
M dwarf mass function is shown in figure 4. As in figure 3, we have plotted
the bins centered on integer magnitudes ({\it open circles}). 
We find the mass function to be essentially flat from the highest mass M 
dwarfs to $M \sim 0.1 M_{\odot}$. Quantitatively, a power-law fit to our 
data over the range $0.6 M_{\odot} > M > 0.1 M_{\odot}$ yields an index 
$\alpha = -0.32 \pm 0.15$; $\alpha = \frac{d log N}{d log M}$. 

A deviation from a Salpeter mass function for $M < 0.6 M_{\odot}$ has been 
previously reported by Miller \& Scalo (1979), Kroupa, Tout, \& Gilmore (1993), 
Reid et al.\ (1995), and Gould et al.\ (1996). Gould et al.\ (1997) find the 
break in the power law to occur at $M = 0.6 M_{\odot}$. At the high mass end, 
they measure a power-law index $\alpha = -1.21$ based on the luminosity 
function of Wielen, Jahreiss, \& Kr\"uger (1983). For $M < 0.6 M_{\odot}$, 
they find $\alpha = 0.44$ to provide the best fit to their HST sample. Reid \& 
Gizis (1997) find the mass function to be flat ($\alpha = -0.05$) over the 
mass range 1 to 0.1 $M_{\odot}$ for stars within 8 pc. 

As stated above, a change in our adopted color-magnitude relation can affect 
the shape of the luminosity function; similarly, any uncertainties in this 
relation could alter the shape of the mass function. We explored the 
direction and magnitude of this effect by varying the color-magnitude 
relation by $1\sigma$ and rederiving the mass function. We found these 
systematic changes to have a negligible effect on the mass function for 
$M > 0.25 M_{\odot}$, while at the low-mass end they can significantly affect 
the steepness of the power-law slope. 

Previous investigators have found evidence for an upturn in the mass 
function near the hydrogen burning limit. We do not 
find any stars with $M_V \geq 16$ and are unable to comment on the shape
of the mass function in this regime. Extrapolating the luminosity function 
of Gould et al.\  (1997) to our sample leads us to expect on order one star 
beyond $M_V \sim 15$. 

\section{Conclusions}

We have used a combination of multicolor photometry and morphological 
information to extend a search for M dwarfs to fainter apparent magnitudes than 
previously possible without the additional observing overhead of spectroscopy. 
This catalog includes 1575 M dwarfs, including 499 in bins over the range 
$12 \leq M_V \leq 15$, and thus 
provides a good statistical sample of these faint stars near the 
hydrogen-burning limit. Our derived luminosity and mass function for M 
dwarfs is consistent with previous work in that it finds a definite 
turnover in the stellar luminosity function fainter than $M_V = 12$. 
For $M_V < 12$ our results are in good agreement with other investigators. 
For stars fainter than the turnover our estimated space densities lie 
between the ground-based photometric parallax values of Reid \& Gizis (1997) 
and the HST results of Gould et al.\ (1997). 
Our mass function is relatively flat for $M < 0.6 M_{\odot}$ and is 
inconsistent with a Salpeter function at the $3\sigma$ level. 
We thus conclude that 
the mass function significantly flattens out in the M dwarf mass regime, 
in broad agreement with the work of Reid \& Gizis (1997) and Gould et al.\
(1997). 

The technique we have presented in this paper can easily be applied to other
large area, ground-based surveys. We have shown that multicolor photometry 
alone can be used to discriminate M dwarfs from galaxies to at least 
$V \sim 22$. Many square degrees could thus be efficiently surveyed with modern
large-format cameras to collect a large sample of the faintest M dwarfs. 
Such a survey would provide the necessary statistics to improve the 
parameterization of the M dwarf density distribution and to accurately 
determine the shape of the mass function near the hydrogen-burning limit.  

\vskip 24pt

\acknowledgments

We would like to thank Andy Gould for several helpful suggestions and comments
on this manuscript. We would also like to thank the referee for 
several valuable suggestions that have improved this paper and our 
presentation. In addition, we would like to thank Pat Hall for providing 
additional information on his work with this survey. This research has
made use of the Simbad database, operated at CDS, Strasbourg, France. This 
work was supported in part by NSF grant AST-95 29324 to PSO.

\newpage

\begin{center}
\begin{tabular}{cccccc}
\multicolumn{4}{c}{{\bf TABLE 1}}\\[12pt]
\multicolumn{4}{c}{{\bf Survey Field Characteristics}}\\[12pt]
\hline
\hline
\multicolumn{1}{c}{Field ID} &
\multicolumn{1}{c}{$l_{II}$} &
\multicolumn{1}{c}{$b_{II}$} &
\multicolumn{1}{c}{A [arcmin$^2$]} \\
\hline
01      & 129 & -63  &  517.3 \\
10      & 248 &  47  &  286.0 \\
14      & 337 &  57  &  561.3 \\
17      & 77  &  35  &  551.9 \\
21      & 52  & -39  &  537.0 \\
22      & 68  & -51  &  536.6 \\
\hline
\end{tabular}
\end{center}

\noindent
Table 1: A summary of the six survey fields of Hall et al. (1996a). Column
1 shows the field identification, which corresponds to the right ascension
of the field in hours. Columns 2 and 3 are the galactic coordinates in
degrees, and column 4 gives the angular size of each field in square
arcminutes.

\begin{center}
\begin{tabular}{ccc}
\multicolumn{3}{c}{{\bf TABLE 2}}\\[12pt]
\multicolumn{3}{c}{{\bf Stellar Density Parameters}}\\[12pt]
\hline
\hline
\multicolumn{1}{c}{Parameter} &
\multicolumn{1}{c}{Value} &
\multicolumn{1}{c}{$\sigma$} \\
\hline
$\alpha$ &  0.79 & 0.12  \\
$h1$     &  340  & 40    \\
$h2$     &  550  & 180   \\
\hline
\end{tabular}
\end{center}

\noindent
Table 2: The parameters for our best-fit, two-component model of the
vertical stellar density distribution. This model contains both a sech$^2$ and
an exponential term. The parameter $\alpha$ is the relative weight
of the sech$^2$ term; the exponential term is weighted by $(1 - \alpha)$.
$h1$ and $h2$ are the scale heights for the sech$^2$ and the exponential
distributions, respectively. The associated uncertainties, $\sigma$,
correspond to the change in a given parameter that would raise the reduced
$\chi^2$ by 1 with the other parameters held fixed. See section 4 for more
details.

\begin{center}
\begin{tabular}{lccc}
\multicolumn{4}{c}{{\bf TABLE 3}}\\[12pt]
\multicolumn{4}{c}{{\bf $V/V_{max}$ Results}}\\[12pt]
\hline
\hline
\multicolumn{1}{c}{$V$ limit} &
\multicolumn{1}{c}{$V/V_{max}$} &
\multicolumn{1}{c}{$\sigma_{V/V_{max}}$} &
\multicolumn{1}{c}{N} \\
\hline
19.0 & 0.4963 & 0.0201 & 207 \nl
19.5 & 0.4660 & 0.0170 & 291 \nl
20.0 & 0.4727 & 0.0142 & 416 \nl
20.5 & 0.5025 & 0.0116 & 618 \nl
21.0 & 0.5136 & 0.0098 & 876 \nl
21.5 & 0.5161 & 0.0084 & 1182 \nl
22.0 & 0.5263 & 0.0073 & 1575 \nl
22.5 & 0.5582 & 0.0062 & 2159 \nl
23.0 & 0.5879 & 0.0053 & 2927 \nl
23.5 & 0.6227 & 0.0046 & 4005 \nl
\hline
\end{tabular}
\end{center}

\noindent
Table 3: A summary of the $V/V_{max}$ calculations described in section 5.
Column 1 lists the limiting $V$ magnitude of each sample. Columns 2 and
3 report the value of $V/V_{max}$ and the $1\sigma$ uncertainties. Column
4 contains the number of stars included in the sample. 

\begin{center}
\begin{tabular}{ccccc}
\multicolumn{5}{c}{{\bf TABLE 4}}\\[12pt]
\multicolumn{5}{c}{{\bf $V/V_{max}$ Results}}\\[12pt]
\hline
\hline
\multicolumn{1}{c}{M$_{V}$} &
\multicolumn{1}{c}{N} &
\multicolumn{1}{c}{stars pc$^{-3}$ ($\times 10^{-3}$)} &
\multicolumn{1}{c}{log[M/M$_{\odot}$]} &
\multicolumn{1}{c}{log[stars pc$^{-3}$ log[M/M$_{\odot}$]$^{-1}$]} \\
\hline
9       & 292   & 4.2 & -0.239 & -1.06  \nl
10      & 415   & 7.8 & -0.289 & -0.98  \nl
11      & 397   & 11.5 & -0.427 & -1.16  \nl
12      & 299   & 14.5 & -0.596 & -1.05 \nl
13      & 126   & 11.6 & -0.755 & -1.00 \nl
14      & 37    & 7.4 & -0.849 & -0.95  \nl
15      & 9     & 4.6 & -0.931 & -1.07  \nl
\hline
\end{tabular}
\end{center}

\noindent
Table 4: The stellar luminosity and mass function derived with this sample.
The stellar density has been normalized to the local neighborhood.
Columns 1 and 2 list the center of each magnitude bin and the corresponding
number of stars in this bin, while column 3 lists the stellar luminosity
function. Column 4 lists log[$M/M_{\odot}$] corresponding to the $M_V$
given in column 1, while column 5 contains the stellar mass function.

\newpage

\begin{figure}
\plotfiddle{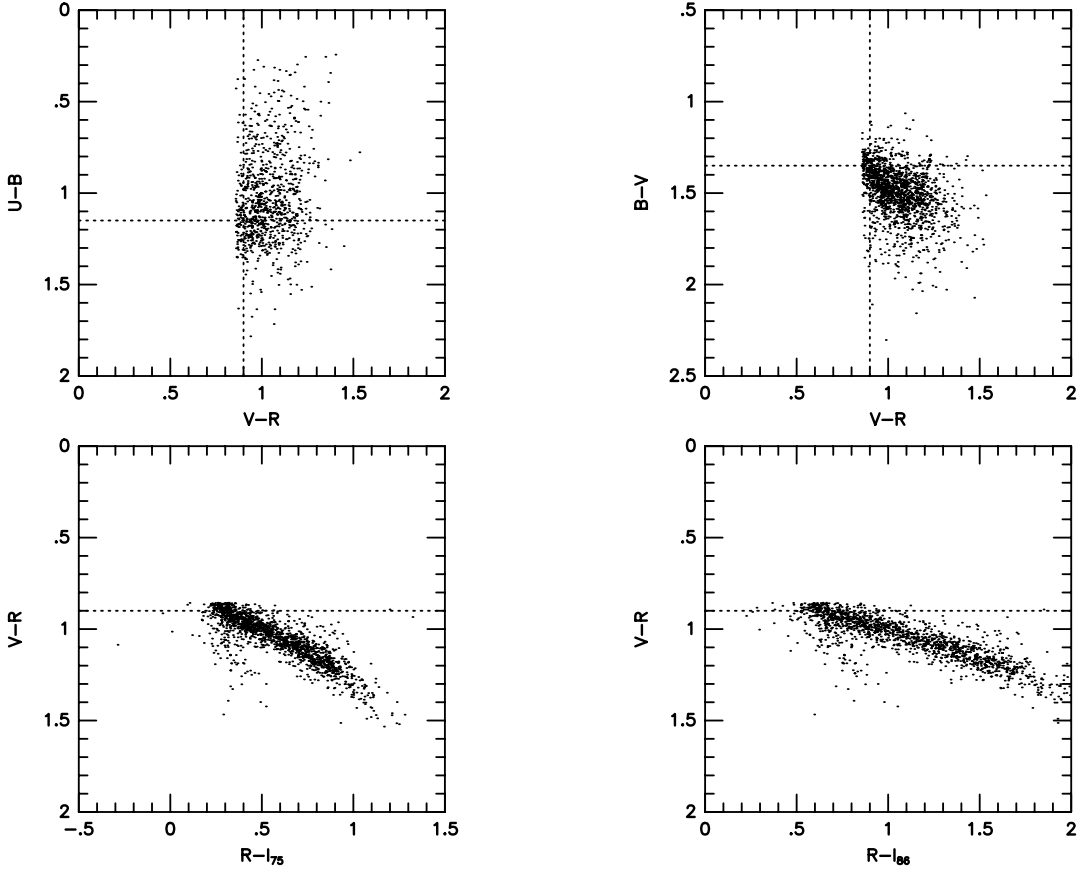}{5.0truein}{0}{70}{70}{-280}{0}
\caption{Color-color diagrams of the M dwarf sample down 
to a limiting $V$ magnitude of 22. The upper left panel shows $U-B$ vs. $V-R$, 
the upper right panel shows $B-V$ vs. $V-R$ and the two lower panels show
$V-R$ vs.\ $R-I_{75}$ and $R-I_{86}$, respectively. Photometric errors 
are generally 0.02, increasing to 0.05 for objects near the magnitude limit.
The dashed lines correspond to the color cut-offs for M dwarfs described in 
the text, with the M dwarfs falling to the lower right in each plot. Stars
outside of this range are consistent with these colors within $3\sigma$.}
\end{figure}

\begin{figure}
\plotfiddle{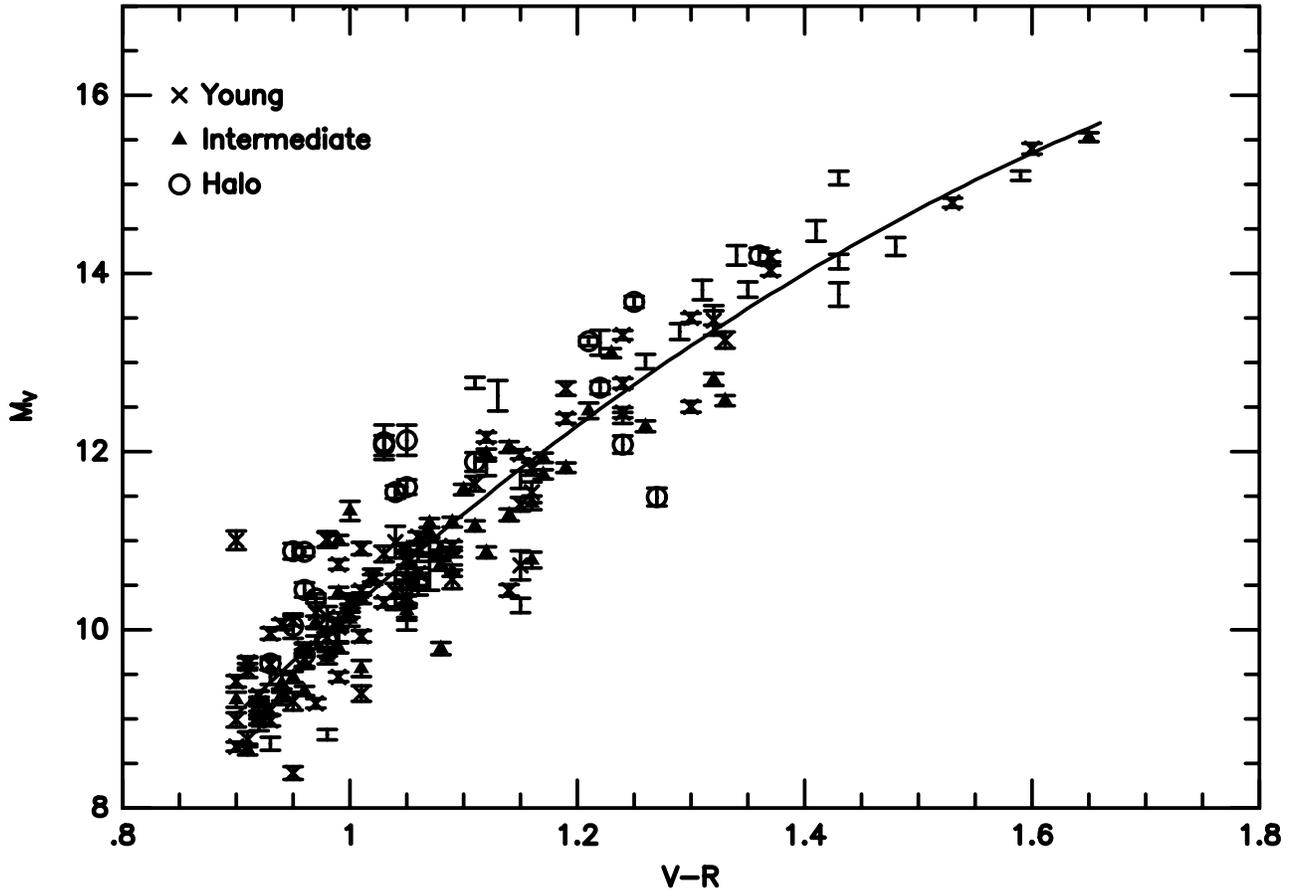}{5.0truein}{0}{70}{70}{-280}{0}
\caption{The color-absolute magnitude relation we derived
for M dwarfs. This M dwarf sample is from the work of Leggett (1992), though 
we include new Hipparcos parallax data when available. M dwarfs that are 
kinematically members of the halo, intermediate, and young disk population 
are plotted as open circles, filled triangles, and crosses, respectively. 
We plot only error bars if the kinematic population type is not available.
This relation is discussed in detail in section 3. }
\end{figure}

\begin{figure}
\plotfiddle{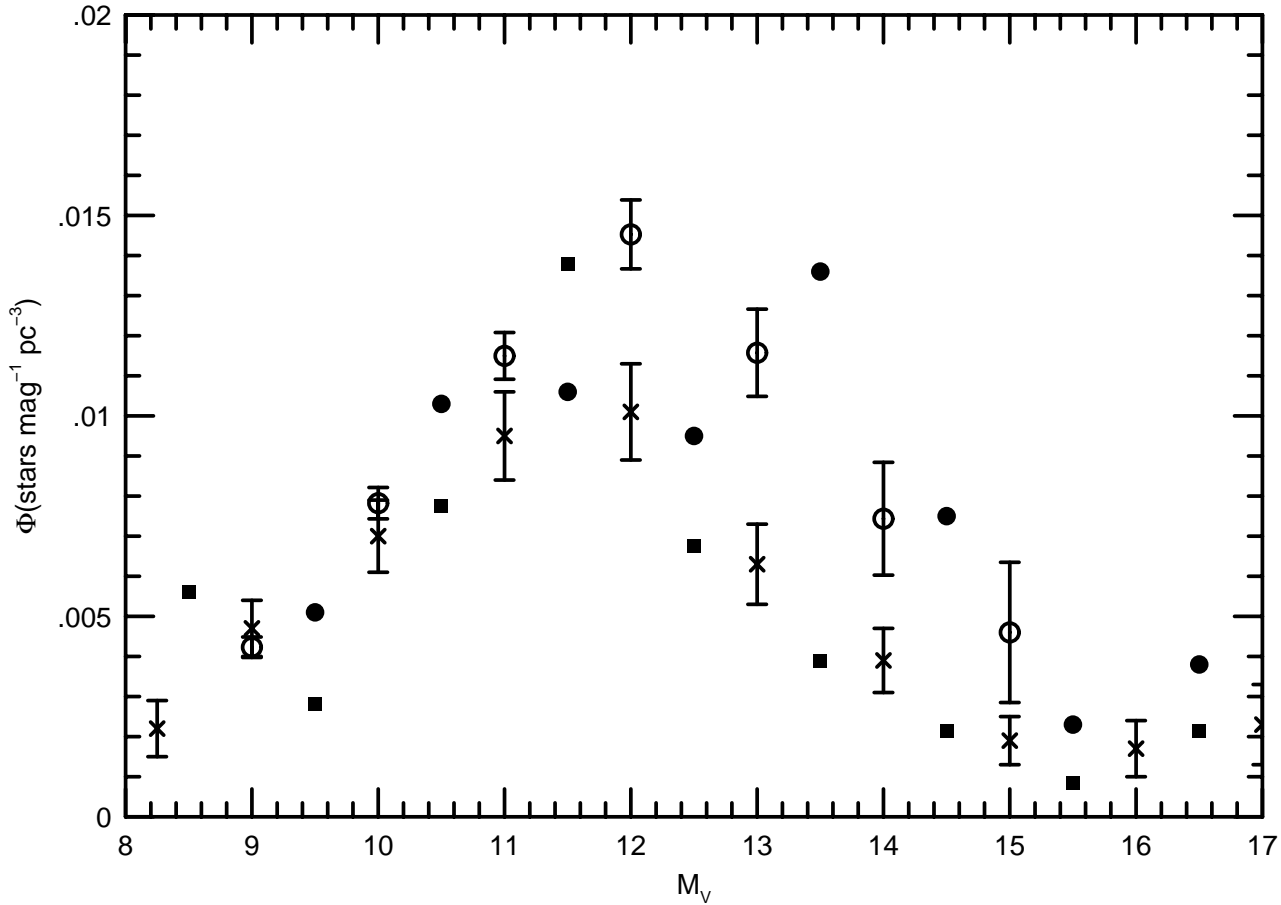}{5.0truein}{0}{70}{70}{-280}{0}
\caption{The stellar luminosity function of M dwarfs at 
the galactic plane as listed in table 4. 
Open circles represent our derived stellar luminosity function; error bars 
are 1$\sigma$ uncertainties based on Poisson statistics. 
For comparison, we also plot the stellar luminosity function derived by 
Stobie et al. (1989; filled squares), the rederived photometric 
parallax luminosity function of Reid \& Gizis (1997; filled circles), 
and the luminosity function of Gould et al. (1997; crosses with error bars). 
The data from Gould et al.\ (1997) is from their derivation based on naive 
binning. }
\end{figure}

\begin{figure}
\plotfiddle{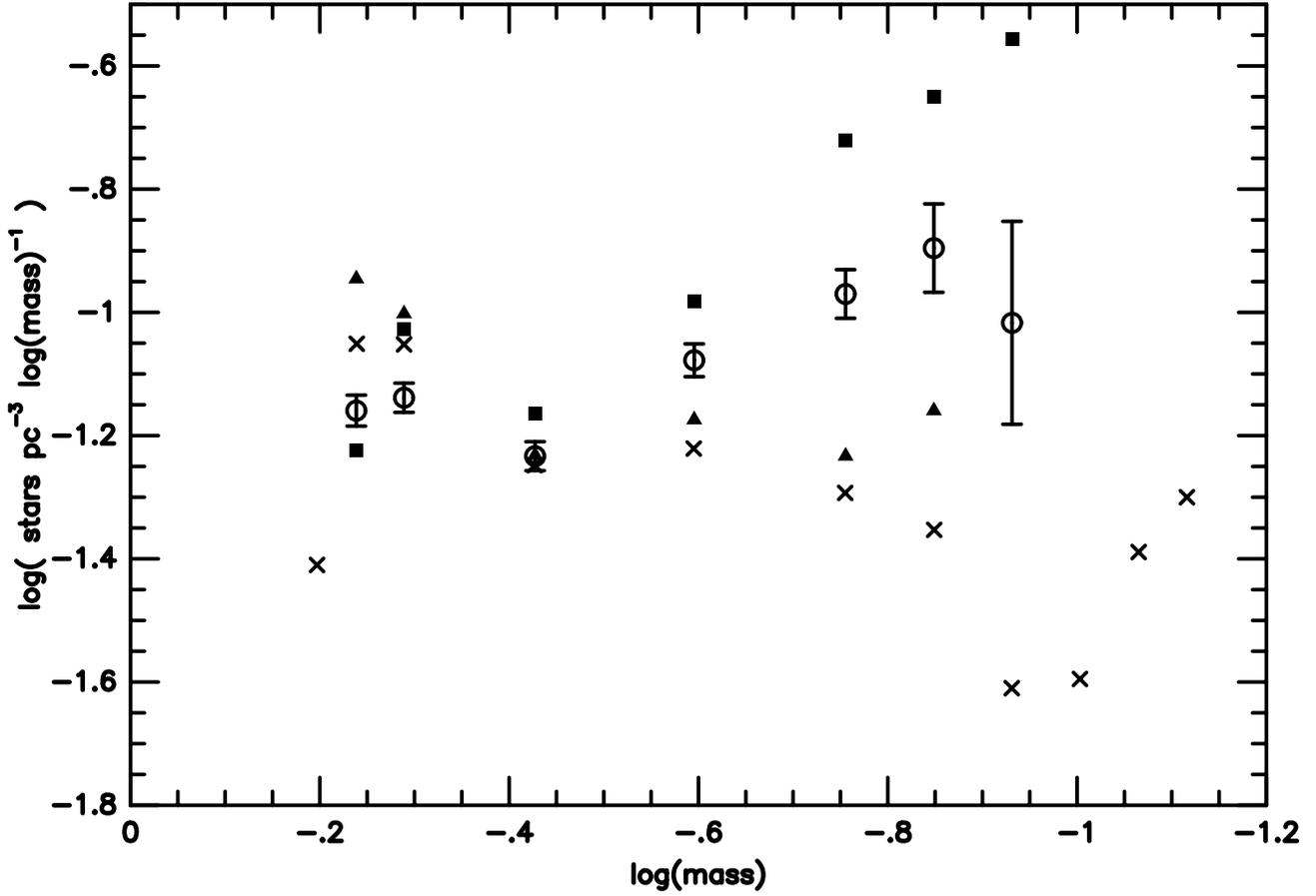}{5.0truein}{0}{70}{70}{-280}{0}
\caption{The M dwarf mass function, derived by employing 
the mass-luminosity 
relation of Henry \& McCarthy (1994). As in figure 2, open circles represent
our derived mass function and crosses represent the mass function derived by 
Gould et al.\ (1997). We also plot the change in our derived mass function if 
we vary our color-magnitude relation by plus (filled squares) and minus 
(filled triangles) $1\sigma$ (see text).}
\end{figure}

\end{document}